\documentstyle[11pt]{article}
\begin{document}
\begin{titlepage}
\hfill{UQMATH-93-01}
\hfill{hep-th/9303095}
\vskip.3in
\begin{center}
{\huge On Universal $R$-Matrix for Quantized Nontwisted Rank $3$ Affine
Lie Algebras}
\vskip.3in
{\Large Y.-Z.Zhang} and {\Large M.D.Gould}
\vskip.3in
{\large Department of Mathematics, University of Queensland, Brisbane,
Qld 4072, Australia}
\end{center}
\vskip.6in
\begin{center}
{\bf Abstract:}
\end{center}
Explicit formulas of the universal $R$-matrix are given for all quantized
nontwisted rank 3 affine Lie algebras $U_q(A_2^{(1)})\,,~U_q(C_2^{(1)})$
and $U_q(G_2^{(1)})$.

\end{titlepage}

\section{Introduction}
To any Kac-Moody (KM) algebra with a symmetrizable, generalized Cartan
matrix in the sense of Kac\cite{Kac} there corresponds a quantum deformation
of its universal enveloping algebra\cite{Drinfeld}\cite{Jimbo}. Such
quantum deformations are found to have a quasitriangular Hopf algebra
structure. In particular, there exists a canonical element $R$ in the deformed
algebra satisfying
the well-known quantum Yang-Baxter relation which plays an
important role in CFT's\cite{Sierra}, quantum integrable models\cite{Faddeev}
\cite{Baxter} and knot theory\cite{Witten}\cite{Reshetikhin}
\cite{ZGB}. The canonical element $R$
is called "universal $R$-matrix".

The explicit form of the $R$-matrix for quantized finite-dimensional simple Lie
algebras  and Lie superalgebras has been known
for some time\cite{Rosso}\cite{KR}\cite{KT1}. Moreover, the general form
of the $R$-matrix for quantum deformations of infinite-dimensional affine
Lie algebras was given in general terms\cite{KT2}. However, the explicit
results
have only been obtained for the quantized nontwisted rank 2 affine Lie algebra.

The aim of the present Letter is to determine the normalizing coeffients
for all quantized nontwisted rank 3 affine Lie algebras. Deriving
these coeffients is interesting in several respects (see section 5),
but is more relevant to our proof on
the unitarity of highest weight modules of quantized affine Lie algebras
which we will publish elsewhere.

\section{Preliminaries}
\noindent
We start with the definition of the nontwisted quantum affine Lie algebra
$U_q({\cal G}^{(1)})$. Let $A^0=(a_{ij})_{1\leq i,j\leq r}$ be a
symmetrizable Cartan matrix. Let ${\cal G}$ stand for the finite-dimensional
simple Lie algebra associated with the symmetrical Cartan matrix
$A^0_{\rm sym}=(a^{\rm sym}_{ij})=(\alpha_i,\alpha_j),~i,j=1,2,...,r$,
where $r$ is the rank of ${\cal G}$.
Let $A=(a_{ij})_{0\leq i,j\leq r}$ be a symmetrizable,
generalized Cartan matrix in the sense of Kac. Let ${\cal G}^{(1)}$ denote
the nontwisted affine Lie algebra associated with the corresponding symmetric
Cartan matrix $A_{\rm sym}=(a^{\rm sym}_{ij})=(\alpha_i,\alpha_j), ~i,j=0,1,
..., r$. Then the
quantum algebra $U_q({\cal G}^{(1)})$ is defined to be a Hopf algebra with
generators: $\{E_i,~F_i,~q^{h_i}~(i=0,1,...,r),~q^d\}$ and relations,
\begin{eqnarray}
&&q^h.q^{h'}=q^{h+h'}~~~~(h,~ h'=h_i~ (i=0,1,...,r),~d)\nonumber\\
&&q^hE_iq^{-h}=q^{(h,\alpha_i)} E_i\,,~~q^hF_iq^{-h}=
  q^{-(h,\alpha_i)}F_i\nonumber\\
&&[E_i, F_j]=\delta_{ij}\frac{q^{h_i}-q^{-h_i}}{q-q^{-1}}\nonumber\\
&&({\rm ad}_qE_i)^{1-a_{ij}}E_j=0\,,~~~({\rm ad}_{q^{-1}}F_i)^{1-a_{ij}}F_j=0
\,~~~~(i\neq j)\label{relations1}
\end{eqnarray}
where
\begin{equation}
({\rm ad}_qx_\alpha)x_\beta=[x_\alpha\,,\,x_\beta]_q=x_\alpha x_\beta -
  q^{(\alpha\,,\,\beta)}x_\beta x_\alpha\nonumber
\end{equation}

The algebra $U_q({\cal G}^{(1)})$ is a Hopf algebra with coproduct, counit and
antipode similar to the case of $U_q(\cal G)$: explicitly, the coproduct is
defined by
\begin{eqnarray}
&&\Delta(q^h)=q^h\otimes q^h\,,~~~h=h_i,~d\nonumber\\
&&\Delta(E_i)=q^{-h_i}\otimes E_i+E_i\otimes 1\nonumber\\
&&\Delta(F_i)=1\otimes F_i+F_i\otimes q^{h_i}\,,~~~i=0,1,...,r
\end{eqnarray}
Formulae for the counit and antipode may also be given, but are not required
below.

Let $\Delta'$ be the opposite coproduct: $\Delta'=T\,\Delta$,~$T(x\otimes
y)=y\otimes x\,,~\forall x,y\in U_q({\cal G}^{(1)})$. Then $\Delta$ and
$\Delta'$ is related by the universal  $R$-matrix $R$
in $U_q({\cal G}^{(1)})\otimes
U_q({\cal G}^{(1)})$ satisfying
\begin{eqnarray}
&&\Delta'(x)R=R\Delta(x)\,,~~~~~~x\in U_q({\cal G}^{(1)})\nonumber\\
&&(\Delta\otimes id )R=R^{13}R^{23}\,,~~~~(id\otimes\Delta)R=R^{13}R^{12}
\end{eqnarray}

We define an anti-involution $\theta$ on
$U_q({\cal G}^{(1)})$ by
\begin{equation}
%&&d^\dagger=d\,,~~h_i^\dagger=h_i\,,~~e_i^\dagger=f_i\,,~~f_i^\dagger=e_i
%  \,,~~~~i=0,1,...,r\nonumber\\
\theta(q^h)=q^{-h}\,,~~\theta(E_i)=F_i\,,~~\theta(F_i)=E_i\,,~~\theta(q)=
  q^{-1}
\end{equation}
which extend uniquely to an algebra anti-involution on
all of $U_q({\cal G}^{(1)})$ so that
$\theta(ab)=\theta(b)\theta(a)\,,~~\forall a,b\in U_q({\cal G}^{(1)})$.
Throughout the paper, we use the notations:
\begin{eqnarray}
&&(n)_q=\frac{1-q^n}{1-q}\,,~~[n]_q=\frac{q^n-q^{-n}}{q-q^{-1}}\,,~~
  q_\alpha=q^{(\alpha,\alpha)}\nonumber\\
&&{\rm exp}_q(x)=\sum_{n\geq 0}\frac{x^n}{(n)_q!}\,,~~(n)_q!=
  (n)_q(n-1)_q\,...\,(1)_q
\end{eqnarray}

\section{Rank $2$ Case: $U_q(A_1^{(1)})$}
This section is devoted to a brief review of Khoroshkin and Tolstoy's
construction\cite{KT2} of the universal $R$-matrix. We start with the
rank 2 case. Fix
a normal ordering in the positive root system $\Delta_+$ of $A_1^{(1)}$ :
\begin{equation}
\alpha,\,\alpha+\delta,\,...,\,\alpha+n\delta,\,...,\,\delta,\,2\delta,\,
...,\,m\delta,\,...\,,\,...\,,\,\beta+l\delta,\,...\,,\beta\label{order1}
\end{equation}
where $\alpha$ and $\beta$ are simple roots; $\delta=\alpha+\beta$ is the
minimal positive imaginary root.
Construct Cartan-Weyl generators $E_\gamma\,,~F_\gamma=\theta(E_\gamma)
\,,~~\gamma\in \Delta_+$ of $U_q(A^{(1)})$ as follows:
We define
\begin{eqnarray}
&&\tilde{E_\delta}=[(\alpha,\alpha)]_q^{-1}[E_\alpha,\,E_\beta]_q\nonumber\\
&&E_{\alpha+n\delta}=(-1)^n\left ({\rm ad}\tilde{E_\delta}\right )^nE_\alpha
  \nonumber\\
&&E_{\beta+n\delta}=\left ({\rm ad}\tilde{E_\delta}\right )^nE_\beta\,,...
  \nonumber\\
&&\tilde{E}_{n\delta}= [(\alpha,\alpha)]_q^{-1}[E_{\alpha+(n-1)\delta},
\,E_\beta]_q  \label{cartan-weyl1}
\end{eqnarray}
where $[\tilde{E}_{n\delta},\,\tilde{E}_{m\delta}]=0$ for any $n,\,m >0$. For
any $n>0$ there exists a unique element $E_{n\delta}$ \cite{KT2} satisfying
$[E_{n\delta}\,,\,E_{m\delta}]=0$ for any $n,\,m>0$ and the relation
\begin{equation}
\tilde{E}_{n\delta}=\sum_{
\begin{array}{c}
k_1p_1+...+k_mp_m=n\\
0<k_1<...<k_m
\end{array}
}\frac{\left ( q^{(\alpha,\alpha)}-q^{-(\alpha,\alpha)}\right )^{\sum_ip_i-1}}
{p_1!\;...\;p_m!}(E_{k_1\delta})^{p_1}...(E_{k_m\delta})^{p_m} \label{ee1}
\end{equation}
Then the vectors $E_\gamma$ and $F_\gamma=
\theta(E_\gamma)$, $\gamma\in \Delta_+$ are the Cartan-Weyl generators for
$U_q(A^{(1)})$.\\
One has\cite{KT2}
\vskip.1in
\noindent {\bf Theorem 3.1:} The universal $R$-matrix for $U_q(A_1^{(1)})
$ may be written as
\begin{eqnarray}
R&=&\left ( \Pi_{n\geq 0}\;{\rm exp}_{q_\alpha}((q-q^{-1})(E_{\alpha+n\delta}
\otimes  F_{\alpha+n\delta}))\right )\nonumber\\
  & &\cdot{\rm exp}\left ( \sum_{n>0}n[n]_{q_\alpha}^{-1}
  (q_\alpha-q_\alpha^{-1})(E_{n\delta}\otimes F_{n\delta})\right )\nonumber\\
& &\cdot\left (\Pi_{n\geq 0}\;{\rm exp}_{q_\alpha}((q-q^{-1})
  (E_{\beta+n\delta}\otimes
  F_{\beta+n\delta}))\right )\cdot
  q^{\frac{1}{2}h_\alpha\otimes h_\alpha+c\otimes d+d\otimes c}\label{sl2R}
\end{eqnarray}
where $c=h_\alpha+h_\beta$. The order in the product (\ref{sl2R}) concides
with the chosen normal order (\ref{order1}).

For the general case $U_q({\cal G}^{(1)})$, KT proposed the following general
construction. Fix some order in the positive
root system $\Delta_+$ of the KM Lie algebra ${\cal G}^{(1)}$, which satisfies
an additional condition,
\begin{equation}
\alpha+n\delta~\leq~k\delta~\leq~(\delta-\beta)+l\delta\label{order2}
\end{equation}
where $\alpha\,,~\beta\,\in~\Delta^0_+\,,~~\Delta_+^0$ is the positive system
of ${\cal G}$\,;~$k\,,\,l\,,\,n\,\geq\,0$ and
$\delta$ is a minimal positive imaginary root.
Then Cartan-Weyl generators $E_\gamma$ and
$F_\gamma=\theta(E_\gamma)\,,~~\gamma
\in \Delta_+$, may be constructed inductively as follows\cite{KT2}.
One starts from simple roots. If, for instance,
$\gamma=\alpha+\beta\,,~~\alpha<\gamma<\beta$, is a root and there are
no other positive roots $\alpha'$ and $\beta'$ between $\alpha$ and $\beta$
such that $\gamma=\alpha'+\beta'$, then set
\begin{equation}
E_\gamma=[E_\alpha\,,\,E_\beta]_q=E_\alpha E_\beta - q^{(\alpha,\beta)}E_\beta
 E_\alpha
\end{equation}
For the root $\delta$, one uses the formula for roots $\alpha+n\delta$
and roots $(\delta-\alpha)+n\delta$, to define
\begin{eqnarray}
&&\tilde{E}_\delta^{(i)}=[(\alpha_i,\alpha_i)]_q^{-1}[E_{\alpha_i},\,
E_{\delta-\alpha_i}]_q\,~~~\alpha_i=\alpha,\,\beta\nonumber\\
&&E_{\alpha_i+n\delta}=(-1)^n\left ({\rm ad}\tilde{E}_\delta^{(i)}\right )^n
E_{\alpha_i}\nonumber\\
&&E_{\delta-\alpha_i+n\delta}=\left ({\rm ad}\tilde{E}_\delta^{(i)}\right )^n
E_{\delta-\alpha_i}\,,~~~~  ...\nonumber\\
&&\tilde{E}_{n\delta}^{(i)}= [(\alpha_i,\alpha_i)]_q^{-1}[E_{\alpha_i
+(n-1)\delta},\,E_{\delta-\alpha_i}]_q  \label{cartan-weyl3}
\end{eqnarray}
The above inductive proceduce may be repeated for other nonsimple roots to
obtain new real root vectors
$E_{\gamma+n\delta}\,,~~E_{\delta-\gamma+n\delta}$\,,
$~\gamma\in\Delta_+^0$.
Finally, the imaginary root vectors $E^{(i)}_{n\delta}$ are defined through
$\tilde{E}^{(i)}_{n\delta}$ by the relation (\ref{ee1}) with $\alpha$
replaced by $\alpha_i$. Then,
the vectors defined above $E^{(i)}_{n\delta}\,,~~
F^{(i)}_{n\delta}=\theta(E^{(i)}_{n\delta})~~(i=1,2)\,,~E_\gamma\,~~
F_\gamma=\theta(E_\gamma)$ are the Cartan-Weyl generators of
$U_q({\cal G}^{(1)})$. KT\cite{KT2} proposed the following:
\vskip.1in
\noindent {\bf Theorem 3.2}: The universal $R$-matrix for
$U_q({\cal G}^{(1)})$ may be written in the form,
\begin{eqnarray}
R&=&\left (\Pi_{\gamma\in \Delta_+^{\rm re}\,,\,\gamma<\delta}~~{\rm exp}_{q_
\gamma}\left (\frac{q-q^{-1}}{C_\gamma(q)}E_\gamma\otimes F_\gamma\right )
\right )\nonumber\\
& &\cdot {\rm exp}\left (\sum_{n>0}\sum^r_{i,j=1}
C^n_{ij}(q)(q-q^{-1})(E^{(i)}_{n\delta}\otimes F^{(j)}_{n\delta})
\right )\nonumber\\
& &\cdot \left (\Pi_{\gamma\in \Delta_+^{\rm re}\,,\,\gamma>\delta}~~
{\rm exp}_{q_
\gamma}\left (\frac{q-q^{-1}}{C_\gamma(q)}E_\gamma\otimes F_\gamma\right )
\right )\cdot q^{\sum^r_{i,j=1}\,(a^{-1}_{\rm sym})^{ij}h_i\otimes h_j
+c\otimes d+d\otimes c}\label{generalR}
\end{eqnarray}
where $c=h_0+h_{\psi}$,~$\psi$ is the highest root of ${\cal G}$,~
$\Delta^{\rm re}_+$ denotes the real root part of $\Delta_+$ and
$(C^n_{ij}(q))=(C^n_{ji}(q))\,,~~i,j=1,2,...,r$, is the inverse of the matrix
$(B^n_{ij}(q))\,~~i,j=1,2,...,r$ with
\begin{equation}
B^n_{ij}(q)=(-1)^{n(1-\delta_{ij})}n^{-1}\frac{q^n_{ij}-q^{-n}_{ij}}
{q_{j}-q^{-1}_{j}}\frac{q-q^{-1}}{q_{i}-q^{-1}_{i}}
\,,~~~~q_{ij}=q^{(\alpha_i,\alpha_j)}\,,~~~q_i\equiv q_{\alpha_i}\label{bij}
\end{equation}
The $C_\gamma(q)$ is a normalizing constant defined by
\begin{equation}
[E_\gamma\,,\,F_\gamma]=\frac{C_\gamma(q)}{q-q^{-1}}\left ( q^{h_\gamma}
-q^{-h_\gamma}\right )
\end{equation}
The order in the product of $R$-matrix coincides with the chosen order
(\ref{order2}).

{}From the point of view of applications, it remains to
compute the normalizing coeffients $C_\gamma\,,~\gamma\in\Delta^{\rm re}_+$
explicitly. This is our task in next section.

\section{Rank $3$ Case: $U_q(A_2^{(1)}),~U_q(C_2^{(1)}),~U_q(G_2^{(1)})$}
In this section we determine the normalizing constants appearing in
(\ref{generalR}) for all quantized nontwisted rank 3 affine Lie algebras.
The rank 3 nontwisted affine Lie algebras are ${\cal G}^{(1)}=A_2^{(1)}\,,
\, C_2^{(1)}\,,\,G_2^{(1)}$.
They correspond to the rank 2 finite-dimensional simple Lie algebras
$A_2\,,\,C_2\,,\,G_2$
with symmetrical Cartan matrix $A^0_{\rm sym}=(a^{\rm sym}_{ij})$,~~
$i,j=1,2$ and positive root system $\Delta^0_+$.
In what follows we use $A^0_{\rm sym}$ in the form
\begin{equation}
A^0_{\rm sym}=(a^{\rm sym}_{ij})=\left (
\begin{array}{cc}
(\alpha,\alpha) & (\alpha,\beta)\\
(\beta,\alpha) & (\beta,\beta)
\end{array}
\right )
\end{equation}
Explicitly,
\begin{equation}
A^0_{\rm sym}=(a^{\rm sym}_{ij})=\left \{
\begin{array}{c}
\left (\begin{array}{cc}
2 & -1\\
-1 & 2
\end{array} \right )\,,~~~~{\rm for}~~ A_2\\
\left (\begin{array}{cc}
2 & -1\\
-1 & 1
\end{array} \right )\,,~~~~{\rm for }~~C_2\\
\left (\begin{array}{cc}
6 & -3\\
-3 & 2
\end{array} \right )\,,~~~~{\rm for}~~G_2
\end{array} \right .
\end{equation}
The simple roots are $\alpha\,,\,\beta$ and $\delta-\psi$ with
$\psi=\alpha+\beta$ for $A_2^{(1)}$,~$\psi=\alpha+2\beta$ for $C_2^{(1)}$
and $\psi=2\alpha+3\beta$ for $G_2^{(1)}$.

Now we come to our main concern, i.e. to determine the normalizing constants
appearing in the above $R$-matrix (\ref{generalR}). First of all, we compute
the inverse of (\ref{bij}) and obtain
\begin{eqnarray}
(C^n_{ij}(q))&=&(C^n_{ji}(q))\nonumber\\
 &=&n\,c_{\alpha,\beta}\,\left (
\begin{array}{cc}
\frac{q^n_\alpha-q^{-n}_\alpha}{q-q^{-1}}(q_\beta-q^{-1}_\beta)^2 &
-(-1)^n\frac{q^n_{\alpha\beta}-q^{-n}_{\alpha\beta}}{q-q^{-1}}
(q_\alpha-q^{-1}_\alpha)
(q_\beta-q^{-1}_\beta)\\
-(-1)^n\frac{q^n_{\alpha\beta}-q^{-n}_{\alpha\beta}}{q-q^{-1}}
(q_\alpha-q^{-1}_\alpha)
(q_\beta-q^{-1}_\beta) & \frac{q_\beta^n-q^{-n}_\beta}{q-q^{-1}}
(q_\alpha-q^{-1}_\alpha)^2
\end{array} \right )
\end{eqnarray}
where
\begin{equation}
c^{-1}_{\alpha\beta}=(q^n_\alpha-q^{-n}_\alpha)(q^n_\beta-q^{-n}_\beta)
-(q^n_{\alpha\beta}-q^{-n}_{\alpha\beta})^2\,,~~~~q_{\alpha\beta}=
q^{(\alpha,\beta)}
\end{equation}
For other normalizing constants we state following
\vskip.1in
\noindent {\bf Proposition 4.1:} For the quantized nontwisted affine algebra
$U_q(A^{(1)}_2)$, we fix the following order in $\Delta_+$ of $A_2^{(1)}$,
\begin{eqnarray}
&&\alpha,\,\alpha+\delta,\,...,\,\alpha+m_1\delta,\,...,\,\alpha+\beta,\,
\alpha+\beta+\delta,\,...,\,\alpha+\beta+m_2\delta,\,...,\,\beta,\,
\beta+\delta,\,...,\,\beta+m_3\delta,\,...,\,\delta,\,
2\delta,\,...,\,\nonumber\\
&&k\delta,\,...,\,...\,(\delta-\beta)+l_1\delta,\,...,\,\delta-\beta,\,...,\,
(\delta-\alpha)+l_2\delta,\,...,\,\delta-\alpha,\,...,\,(\delta-\alpha-\beta)
+l_3\delta,\,...,\,\delta-\alpha-\beta\nonumber\\
\label{ordering1}
\end{eqnarray}
where $m_i,k,l_i \geq 0\,,~~i=1,2,3$. We set
\begin{eqnarray}
&&E_{\alpha+\beta}=[E_\alpha\,,\,E_\beta]_q\,,~~~~~
  E_{\delta-\alpha}=[E_\beta\,,\,E_{\delta-\alpha-\beta}]_q\nonumber\\
&&E_{\delta-\beta}=[E_\alpha\,,\,E_{\delta-\alpha-\beta}]_q
\end{eqnarray}
and use formula (\ref{cartan-weyl3}) for $E_{\gamma+n\delta}$ and
$E_{(\delta-\gamma)+n\delta}$, ~$\gamma\in \Delta_+^0$. Then
the root vectors $E_\gamma\,,F_\gamma=\theta(E_\gamma)\,,\gamma\in
\Delta_+^{\rm re}$ satisfy the following relations:
\begin{eqnarray}
&&[E_{\alpha+n\delta},F_{\alpha+n\delta}]=([(\alpha+\beta,\beta)]_q)^n\cdot
  \frac{q^{h_{\alpha+n\delta}}-q^{-h_{\alpha+n\delta}}}{q-q^{-1}}\nonumber\\
&&[E_{\beta+n\delta},F_{\beta+n\delta}]=([(\alpha+\beta,\alpha)]_q)^n\cdot
  \frac{q^{h_{\beta+n\delta}}-q^{-h_{\beta+n\delta}}}{q-q^{-1}}\nonumber\\
&&[E_{(\delta-\alpha-\beta)+n\delta},F_{(\delta-\alpha-\beta)+n\delta}]=
(-[(\alpha,\beta)]_q)^n\cdot
  \frac{q^{h_{(\delta-\alpha-\beta)+n\delta}}-q^{-h_{(\delta-\alpha-\beta)
  +n\delta}}}{q-q^{-1}}\nonumber\\
&&[E_{\alpha+\beta+n\delta},F_{\alpha+\beta+n\delta}]=
(-[(\alpha,\beta)]_q)^{n+1}\cdot
\frac{q^{h_{\alpha+\beta+n\delta}}-q^{-h_{\alpha+\beta+n\delta}}}{q-q^{-1}}
  \nonumber\\
&&[E_{(\delta-\alpha)+n\delta},F_{(\delta-\alpha)+n\delta}]=
([(\alpha+\beta,\beta)]_q)^{n+1}\cdot
  \frac{q^{h_{(\delta-\alpha)+n\delta}}-q^{-h_{(\delta-\alpha)+n\delta}}}
  {q-q^{-1}}\nonumber\\
&&[E_{(\delta-\beta)+n\delta},F_{(\delta-\beta)+n\delta}]=
([(\alpha+\beta,\alpha)]_q)^{n+1}\cdot
\frac{q^{h_{(\delta-\beta)+n\delta}}-q^{-h_{(\delta-\beta)+n\delta}}}
{q-q^{-1}}\label{a2}
\end{eqnarray}
\vskip.1in
\noindent {\bf Proposition 4.2:} For the quantized nontwisted affine algebra
$U_q(C_2^{(1)})$, we fix the order in $\Delta_+$ of $C_2^{(1)}$, according to
\begin{eqnarray}
&&\alpha,\,\alpha+\delta,\,...,\,\alpha+m_1\delta,\,...,\,\alpha+\beta,\,
\alpha+\beta+\delta,\,...,\,\alpha+\beta+m_2\delta,\,...,\,\alpha+2\beta,\,
\alpha+2\beta+\delta,\,...,\,\nonumber\\
&&\alpha+2\beta+m_3\delta,\,...,\,
\beta,\,\beta+\delta,\,...,\,
\beta+m_4\delta,\,...,\,\delta,\,2\delta,\,...,\,
k\delta,\,...,\,...\,(\delta-\beta)+l_1\delta,\,...,\,\delta-\beta,\,...,\,
\nonumber\\
&&(\delta-\alpha)+l_2\delta,\,...,\,\delta-\alpha,\,...,\,(\delta-\alpha-\beta)
+l_3\delta,\,...,\,\delta-\alpha-\beta,\,...,\,(\delta-\alpha-2\beta)+l_4
\delta,\,...,\,\delta-\alpha-2\beta\nonumber\\
\label{ordering2}
\end{eqnarray}
where $m_i,k,l_i\geq 0\,,~~i=1,2,3,4$. We set
\begin{eqnarray}
&&E_{\alpha+\beta}=[E_\alpha\,,\,E_\beta]_q\,,~~~~~
  E_{\alpha+2\beta}=[E_{\alpha+\beta}, E_\beta]_q\nonumber\\
&&E_{\delta-\alpha}=[E_\beta, E_{\delta-\alpha-\beta}]_q\,,~~~~~
  E_{\delta-\beta}=[E_\alpha, E_{\delta-\alpha-\beta}]_q\nonumber\\
&&E_{\delta-\alpha-\beta}=[E_\beta, E_{\delta-\alpha-2\beta}]_q
\end{eqnarray}
and use formula (\ref{cartan-weyl3}) for $E_{\gamma+n\delta}$ and
$E_{(\delta-\gamma)+n\delta}$, ~$\gamma\in \Delta_+^0$. Then
the root vectors $E_\gamma\,,F_\gamma=\theta(E_\gamma)\,,\gamma\in
\Delta_+^{\rm re}$ satisfy the following relations
\begin{eqnarray}
&&[E_{\alpha+n\delta},F_{\alpha+n\delta}]=([(\alpha+2\beta,\beta)]_q)^{2n}\cdot
  \frac{q^{h_{\alpha+n\delta}}-q^{-h_{\alpha+n\delta}}}{q-q^{-1}}\nonumber\\
&&[E_{\beta+n\delta},F_{\beta+n\delta}]=([(\alpha+2\beta,\beta)]_q)^n\cdot
 ([(\alpha+\beta,\alpha)]_q)^n\cdot
  \frac{q^{h_{\beta+n\delta}}-q^{-h_{\beta+n\delta}}}{q-q^{-1}}\nonumber\\
&&[E_{(\delta-\alpha-2\beta)+n\delta},F_{(\delta-\alpha-2\beta)+n\delta}]=
([(\alpha,\beta)]_q)^{2n}
\cdot\frac{q^{h_{(\delta-\alpha-2\beta)+n\delta}}-
q^{-h_{(\delta-\alpha-2\beta)+n\delta}}}{q-q^{-1}}\nonumber\\
&&[E_{\alpha+\beta+n\delta},F_{\alpha+\beta+n\delta}]=
(-[(\alpha,\beta)]_q)^{n+1}\cdot
  \frac{q^{h_{\alpha+\beta+n\delta}}-q^{-h_{\alpha+\beta+n\delta}}}{q-q^{-1}}
  \nonumber\\
&&[E_{\alpha+2\beta+n\delta},F_{\alpha+2\beta+n\delta}]=
([(\alpha,\beta)]_q)^{2(n+1)}\cdot
 \frac{q^{h_{\alpha+2\beta+n\delta}}-q^{-h_{\alpha+2\beta+n\delta}}}{q-q^{-1}}
  \nonumber\\
&&[E_{(\delta-\alpha-\beta)+n\delta},F_{(\delta-\alpha-\beta)+n\delta}]=
([(\alpha+2\beta,\beta)]_q)^{n+1}\cdot
  \frac{q^{h_{(\delta-\alpha-\beta)+n\delta}}-q^{-h_{(\delta-\alpha-\beta)
 +n\delta}}}{q-q^{-1}}\nonumber\\
&&[E_{(\delta-\alpha)+n\delta},F_{(\delta-\alpha)+n\delta}]=
([(\alpha+2\beta,\beta)]_q)^{2(n+1)}\cdot
  \frac{q^{h_{(\delta-\alpha)+n\delta}}-q^{-h_{(\delta-\alpha)+n\delta}}}
  {q-q^{-1}}\nonumber\\
&&[E_{(\delta-\beta)+n\delta},F_{(\delta-\beta)+n\delta}]=
([(\alpha+2\beta,\beta)]_q)^{n+1}
 \cdot ([(\alpha+\beta,\alpha)]_q)^{n+1}
\cdot\frac{q^{h_{(\delta-\beta)+n\delta}}-q^{-h_{(\delta-\beta)+n\delta}}}
 {q-q^{-1}}\nonumber\\
 \label{c2}
\end{eqnarray}
\vskip.1in
\noindent {\bf Proposition 4.3:} For the quantized nontwisted affine algebra
$U_q(G_2^{(1)})$, we fix the order in $\Delta_+$ of $G_2^{(1)}$ as follows:
\begin{eqnarray}
&&\alpha,\,\alpha+\delta,\,...,\,\alpha+m_1\delta,\,...,\,\alpha+\beta,\,
\alpha+\beta+\delta,\,...,\,\alpha+\beta+m_2\delta,\,...,\,2\alpha+3\beta,\,
2\alpha+3\beta+\delta,\,...,\,\nonumber\\
&&2\alpha+3\beta+m_3\delta,\,...,\,\alpha+2\beta,\,
\alpha+2\beta+\delta,\,...,\,\alpha+2\beta+m_4\delta,\,...,\,
\alpha+3\beta,\,\alpha+3\beta+\delta,\,...,\,\nonumber\\
&&\alpha+3\beta+m_5\delta,\,...,\,\beta,\,\beta+\delta,\,...,\,
\beta+m_6\delta,\,...,\,\delta,\,2\delta,\,...,\,
k\delta,\,...,\,...\,(\delta-\beta)+l_1\delta,\,...,\,\delta-\beta,\,...,\,
\nonumber\\
&&(\delta-\alpha)+l_2\delta,\,...,\,\delta-\alpha,\,...,\,
(\delta-\alpha-\beta)+l_3\delta,\,...,\,\delta-\alpha-\beta,\,...,\,
(\delta-\alpha-2\beta)+l_4\delta,\,...,\,\nonumber\\
&&\delta-\alpha-2\beta,\,...,\,(\delta-\alpha-3\beta)+l_5
\delta,\,...,\,\delta-\alpha-3\beta,\,...,\,(\delta-2\alpha-3\beta)+l_6
\delta,\,...,\,\delta-2\alpha-3\beta\nonumber\\
\label{ordering3}
\end{eqnarray}
where $m_i,k,l_i\geq 0\,,~~i=1,2,...,6$. We set
\begin{eqnarray}
&&E_{\alpha+\beta}=[E_\alpha,E_\beta]_q\,,~~~~~E_{\alpha+2\beta}=[E_{\alpha+
  \beta},E_\beta]_q\nonumber\\
&&E_{\alpha+3\beta}=[E_{\alpha+2\beta},E_\beta]_q\,,~~~~~E_{2\alpha+3\beta}=
  [E_{\alpha+\beta},E_{\alpha+2\beta}]_q\nonumber\\
&&E_{\delta-\alpha-3\beta}=[E_\alpha, E_{\delta-2\alpha-3\beta}]_q\,,~~~~~
  E_{\delta-\alpha-2\beta}=[E_\beta, E_{\delta-\alpha-3\beta}]_q\nonumber\\
&&E_{\delta-\alpha-\beta}=[E_\beta, E_{\delta-\alpha-2\beta}]_q\,,~~~~~
  E_{\delta-\alpha}=[E_\beta, E_{\delta-\alpha-\beta}]_q\nonumber\\
&&E_{\delta-\beta}=[E_\alpha, E_{\delta-\alpha-\beta}]_q
\end{eqnarray}
and use formula (\ref{cartan-weyl3}) for $E_{\gamma+n\delta}$ and
$E_{(\delta-\gamma)+n\delta}$, ~$\gamma\in \Delta_+^0$. Then
the root vectors $E_\gamma\,,F_\gamma=\theta(E_\gamma)\,,\gamma\in
\Delta_+^{\rm re}$ satisfy the following relations
\begin{eqnarray}
&&[E_{\alpha+n\delta},
F_{\alpha+n\delta}]=a^n\cdot([(\alpha+3\beta,\beta)]_q)^n

\cdot\frac{q^{h_{\alpha+n\delta}}-q^{-h_{\alpha+n\delta}}}{q-q^{-1}}\nonumber\\
&&[E_{\beta+n\delta}, F_{\beta+n\delta}]=a^n\cdot (
  [(\alpha+\beta,\alpha)]_q)^n\cdot
  \frac{q^{h_{\beta+n\delta}}-q^{-h_{\beta+n\delta}}}{q-q^{-1}}\nonumber\\
&&[E_{(\delta-2\alpha-3\beta)+n\delta},
F_{(\delta-2\alpha-3\beta)+n\delta}]=b^n
  \cdot([(\alpha,\beta)]_q)^{2n}
  \cdot\frac{q^{h_{(\delta-2\alpha-3\beta)+n\delta}}-q^{-h_{(\delta-2\alpha
 -3\beta)+n\delta}}}{q-q^{-1}}\nonumber\\
&&[E_{\alpha+\beta+n\delta}, F_{\alpha+\beta+n\delta}]=(-[(\alpha,\beta)
  ]_q)^{n+1}\cdot\frac{q^{h_{\alpha+\beta+n\delta}}-q^{-h_{\alpha+\beta
  +n\delta}}}{q-q^{-1}}\nonumber\\
&&[E_{\alpha+2\beta+n\delta}, F_{\alpha+2\beta+n\delta}]=b^{n+1}\cdot
  \frac{q^{h_{\alpha+2\beta+n\delta}}-q^{-h_{\alpha+2\beta
  +n\delta}}}{q-q^{-1}}\nonumber\\
&&[E_{\alpha+3\beta+n\delta}, F_{\alpha+3\beta+n\delta}]=b^{n+1}\cdot
  (-[(\alpha,\beta)]_q)^{n+1}\cdot
  \frac{q^{h_{\alpha+3\beta+n\delta}}-q^{-h_{\alpha+3\beta
  +n\delta}}}{q-q^{-1}}\nonumber\\
&&[E_{2\alpha+3\beta+n\delta}, F_{2\alpha+3\beta+n\delta}]=b^{n+1}\cdot
  ([(\alpha,\beta)])^{2(n+1)}\cdot
  \frac{q^{h_{2\alpha+3\beta+n\delta}}-q^{-h_{2\alpha+3\beta
  +n\delta}}}{q-q^{-1}}\nonumber\\
&&[E_{(\delta-\alpha-3\beta)+n\delta}, F_{(\delta-\alpha-3\beta)+n\delta}]=
 ([(2\alpha+3\beta,\alpha)]_q)^{n+1}\cdot
  \frac{q^{h_{(\delta-\alpha-3\beta)+n\delta}}-q^{-h_{(\delta-\alpha-3\beta)
  +n\delta}}}{q-q^{-1}}\nonumber\\
&&[E_{(\delta-\alpha-2\beta)+n\delta}, F_{(\delta-\alpha-2\beta)+n\delta}]=
  ([(2\alpha+3\beta,\alpha)]_q[(\alpha+3\beta,\beta)]_q )^{n+1}\cdot
  \frac{q^{h_{(\delta-\alpha-2\beta)+n\delta}}-q^{-h_{(\delta-\alpha-2\beta)
  +n\delta}}}{q-q^{-1}}\nonumber\\
&&[E_{(\delta-\alpha-\beta)+n\delta},
F_{(\delta-\alpha-\beta)+n\delta}]=a^{n+1}
 \cdot \frac{q^{h_{(\delta-\alpha-\beta)+n\delta}}-q^{-h_{(\delta-\alpha-\beta)
  +n\delta}}}{q-q^{-1}}\nonumber\\
&&[E_{(\delta-\alpha)+n\delta}, F_{(\delta-\alpha)+n\delta}]=a^{n+1}\cdot (
  [(\alpha+3\beta,\beta)]_q)^{n+1}\cdot
  \frac{q^{h_{(\delta-\alpha)+n\delta}}-q^{-h_{(\delta-\alpha)
  +n\delta}}}{q-q^{-1}}\nonumber\\
&&[E_{(\delta-\beta)+n\delta}, F_{(\delta-\beta)+n\delta}]=a^{n+1}\cdot
  ([(\alpha+\beta,\alpha)]_q)^{n+1}\cdot
  \frac{q^{h_{(\delta-\beta)+n\delta}}-q^{-h_{(\delta-\beta)
  +n\delta}}}{q-q^{-1}}\label{g2}
\end{eqnarray}
where
\begin{eqnarray}
&&a=[(2\alpha+3\beta,\alpha)]_q\cdot [(\alpha+3\beta,\beta)]_q\cdot
  ([\alpha+3\beta,\beta)]_q+[(\alpha+2\beta,\beta)]_q)\nonumber\\
&&b=[(\alpha,\beta)]_q\cdot
  ([(\alpha,\beta)]_q+[\alpha+\beta,\beta)]_q)
\end{eqnarray}
\vskip.1in
\noindent{\bf Proof:} All these propositions 4.1--4.3 can be obtained by
direct computations and induction on $n$.~~~$\Box$

Now that all explicit expressions for the constants $C_\gamma$ appearing the
$R$-matrix have been obtained, we deduce, from theorem 3.2,
\vskip.1in
\noindent{\bf Theorem 4.1:} For $U_q(A_2^{(1)})$, the universal $R$-matrix
takes the explicit form
\begin{eqnarray}
R&=&\left (\Pi_{n\geq 0}~{\rm exp}_{q_\alpha}
\left (\frac{q-q^{-1}}{C_{\alpha+n\delta}(q)}E_{\alpha+n\delta}
\otimes F_{\alpha+n\delta}\right )\right )\nonumber\\
& &\cdot\left (\Pi_{n\geq 0}~{\rm exp}_{q_{\alpha+\beta}}
\left (\frac{q-q^{-1}}{C_{\alpha+\beta+n\delta}(q)}E_{\alpha+\beta+n\delta}
\otimes F_{\alpha+\beta+n\delta}\right )\right )\nonumber\\
& &\cdot \left (\Pi_{n\geq 0}~{\rm exp}_{q_\beta}
\left (\frac{q-q^{-1}}{C_{\beta+n\delta}(q)}E_{\beta+n\delta}
\otimes F_{\beta+n\delta}\right )\right )\nonumber\\
& &\cdot {\rm exp}\left (\sum_{n>0}\sum^2_{i,j=1}
C^n_{ij}(q)(q-q^{-1})(E^{(i)}_{n\delta}\otimes F^{(j)}_{n\delta})
\right )\nonumber\\
& &\cdot \left (\Pi_{n\geq 0}~{\rm exp}_{q_{\beta}}
\left (\frac{q-q^{-1}}{C_{(\delta-\beta)+n\delta}(q)}E_{(\delta-\beta)+n\delta}
\otimes F_{(\delta-\beta)+n\delta}\right )\right )\nonumber\\
& &\cdot \left (\Pi_{n\geq 0}~{\rm exp}_{q_\alpha}
\left (\frac{q-q^{-1}}{C_{(\delta-\alpha)+n\delta}(q)}E_{(\delta-\alpha)
+n\delta}
\otimes F_{(\delta-\alpha)+n\delta}\right )\right )\nonumber\\
& &\cdot\left (\Pi_{n\geq 0}~{\rm exp}_{q_{\alpha+\beta}}
\left (\frac{q-q^{-1}}{C_{(\delta-\alpha-\beta)+n\delta}(q)}
E_{(\delta-\alpha-\beta)+n\delta}
\otimes F_{(\delta-\alpha-\beta)+n\delta}\right )\right )\nonumber\\
& &\cdot q^{\sum^2_{i,j=1}\,(a^{-1}_{\rm sym})^{ij}h_i\otimes h_j+c\otimes d+
d\otimes c}\label{aR}
\end{eqnarray}
where $C_{\gamma+n\delta}\,~C_{(\delta-\gamma)+n\delta}\,,~\gamma\in\Delta^0
_+$ can be read off from (\ref{a2}) and the order in the product of (\ref{aR})
is defined by (\ref{ordering1}).
\vskip.1in
\noindent{\bf Theorem 4.2:} For $U_q(C_2^{(1)})$, the universal $R$-matrix
takes the explicit form
\begin{eqnarray}
R&=&\left (\Pi_{n\geq 0}~{\rm exp}_{q_\alpha}
\left (\frac{q-q^{-1}}{C_{\alpha+n\delta}(q)}E_{\alpha+n\delta}
\otimes F_{\alpha+n\delta}\right )\right )\nonumber\\
& &\cdot \left (\Pi_{n\geq 0}~{\rm exp}_{q_{\alpha+\beta}}
\left (\frac{q-q^{-1}}{C_{\alpha+\beta+n\delta}(q)}E_{\alpha+\beta+n\delta}
\otimes F_{\alpha+\beta+n\delta}\right )\right )\nonumber\\
& &\cdot \left (\Pi_{n\geq 0}~{\rm exp}_{q_{\alpha+2\beta}}
\left (\frac{q-q^{-1}}{C_{\alpha+2\beta+n\delta}(q)}E_{\alpha+2\beta+n\delta}
\otimes F_{\alpha+2\beta+n\delta}\right )\right )\nonumber\\
& &\cdot\left (\Pi_{n\geq 0}~{\rm exp}_{q_\beta}
\left (\frac{q-q^{-1}}{C_{\beta+n\delta}(q)}E_{\beta+n\delta}
\otimes F_{\beta+n\delta}\right )\right )\nonumber\\
& &\cdot {\rm exp}\left (\sum_{n>0}\sum^2_{i,j=1}
C^n_{ij}(q)(q-q^{-1})(E^{(i)}_{n\delta}\otimes F^{(j)}_{n\delta})
\right )\nonumber\\
& &\cdot \left (\Pi_{n\geq 0}~{\rm exp}_{q_{\beta}}
\left (\frac{q-q^{-1}}{C_{(\delta-\beta)+n\delta}(q)}E_{(\delta-\beta)+n\delta}
\otimes F_{(\delta-\beta)+n\delta}\right )\right )\nonumber\\
& &\cdot\left (\Pi_{n\geq 0}~{\rm exp}_{q_\alpha}
\left (\frac{q-q^{-1}}{C_{(\delta-\alpha)+n\delta}(q)}E_{(\delta-\alpha)
+n\delta}
\otimes F_{(\delta-\alpha)+n\delta}\right )\right )\nonumber\\
& &\cdot\left (\Pi_{n\geq 0}~{\rm exp}_{q_{\alpha+\beta}}
\left (\frac{q-q^{-1}}{C_{(\delta-\alpha-\beta)+n\delta}(q)}
E_{(\delta-\alpha-\beta)+n\delta}
\otimes F_{(\delta-\alpha-\beta)+n\delta}\right )\right )\nonumber\\
& &\cdot \left (\Pi_{n\geq 0}~{\rm exp}_{q_{\alpha+2\beta}}
\left (\frac{q-q^{-1}}{C_{(\delta-\alpha-2\beta)+n\delta}(q)}
E_{(\delta-\alpha-2\beta)+n\delta}
\otimes F_{(\delta-\alpha-2\beta)+n\delta}\right )\right )\nonumber\\
& &\cdot q^{\sum^2_{i,j=1}\,(a^{-1}_{\rm sym})^{ij}h_i\otimes h_j+c\otimes d+
d\otimes c}\label{cR}
\end{eqnarray}
where $C_{\gamma+n\delta}\,~C_{(\delta-\gamma)+n\delta}\,,~\gamma\in\Delta^0
_+$ can be read off from (\ref{c2}) and the order in the product of (\ref{cR})
is defined by (\ref{ordering2}).
\vskip.1in
\noindent{\bf Theorem 4.3:} For $U_q(G_2^{(1)})$, the universal $R$-matrix
takes the explicit form
\begin{eqnarray}
R&=&\left (\Pi_{n\geq 0}~{\rm exp}_{q_\alpha}
\left (\frac{q-q^{-1}}{C_{\alpha+n\delta}(q)}E_{\alpha+n\delta}
\otimes F_{\alpha+n\delta}\right )\right )\nonumber\\
& &\cdot\left (\Pi_{n\geq 0}~{\rm exp}_{q_{\alpha+\beta}}
\left (\frac{q-q^{-1}}{C_{\alpha+\beta+n\delta}(q)}E_{\alpha+\beta+n\delta}
\otimes F_{\alpha+\beta+n\delta}\right )\right )\nonumber\\
& &\cdot \left (\Pi_{n\geq 0}~{\rm exp}_{q_{2\alpha+3\beta}}
\left (\frac{q-q^{-1}}{C_{2\alpha+3\beta+n\delta}(q)}E_{2\alpha+3\beta+n\delta}
\otimes F_{2\alpha+3\beta+n\delta}\right )\right )\nonumber\\
& &\cdot\left (\Pi_{n\geq 0}~{\rm exp}_{q_{\alpha+2\beta}}
\left (\frac{q-q^{-1}}{C_{\alpha+2\beta+n\delta}(q)}E_{\alpha+2\beta+n\delta}
\otimes F_{\alpha+2\beta+n\delta}\right )\right )\nonumber\\
& &\cdot\left (\Pi_{n\geq 0}~{\rm exp}_{q_{\alpha+3\beta}}
\left (\frac{q-q^{-1}}{C_{\alpha+3\beta+n\delta}(q)}E_{\alpha+3\beta+n\delta}
\otimes F_{\alpha+3\beta+n\delta}\right )\right )\nonumber\\
& &\cdot\left (\Pi_{n\geq 0}~{\rm exp}_{q_\beta}
\left (\frac{q-q^{-1}}{C_{\beta+n\delta}(q)}E_{\beta+n\delta}
\otimes F_{\beta+n\delta}\right )\right )\nonumber\\
& &\cdot {\rm exp}\left (\sum_{n>0}\sum^2_{i,j=1}
C^n_{ij}(q)(q-q^{-1})(E^{(i)}_{n\delta}\otimes F^{(j)}_{n\delta})
\right )\nonumber\\
& &\cdot\left (\Pi_{n\geq 0}~{\rm exp}_{q_{\beta}}
\left (\frac{q-q^{-1}}{C_{(\delta-\beta)+n\delta}(q)}E_{(\delta-\beta)+n\delta}
\otimes F_{(\delta-\beta)+n\delta}\right )\right )\nonumber\\
& &\cdot\left (\Pi_{n\geq 0}~{\rm exp}_{q_\alpha}
\left (\frac{q-q^{-1}}{C_{(\delta-\alpha)+n\delta}(q)}E_{(\delta-\alpha)
+n\delta}
\otimes F_{(\delta-\alpha)+n\delta}\right )\right )\nonumber\\
& &\cdot\left (\Pi_{n\geq 0}~{\rm exp}_{q_{\alpha+\beta}}
\left (\frac{q-q^{-1}}{C_{(\delta-\alpha-\beta)+n\delta}(q)}
E_{(\delta-\alpha-\beta)+n\delta}
\otimes F_{(\delta-\alpha-\beta)+n\delta}\right )\right )\nonumber\\
& &\cdot \left (\Pi_{n\geq 0}~{\rm exp}_{q_{\alpha+2\beta}}
\left (\frac{q-q^{-1}}{C_{(\delta-\alpha-2\beta)+n\delta}(q)}
E_{(\delta-\alpha-2\beta)+n\delta}
\otimes F_{(\delta-\alpha-2\beta)+n\delta}\right )\right )\nonumber\\
& &\cdot\left (\Pi_{n\geq 0}~{\rm exp}_{q_{\alpha+3\beta}}
\left (\frac{q-q^{-1}}{C_{(\delta-\alpha-3\beta)+n\delta}(q)}
E_{(\delta-\alpha-3\beta)+n\delta}
\otimes F_{(\delta-\alpha-3\beta)+n\delta}\right )\right )\nonumber\\
& &\cdot\left (\Pi_{n\geq 0}~{\rm exp}_{q_{2\alpha+3\beta}}
\left (\frac{q-q^{-1}}{C_{(\delta-2\alpha-3\beta)+n\delta}(q)}
E_{(\delta-2\alpha-3\beta)+n\delta}
\otimes F_{(\delta-2\alpha-3\beta)+n\delta}\right )\right )\nonumber\\
& &\cdot q^{\sum^2_{i,j=1}\,(a^{-1}_{\rm sym})^{ij}h_i\otimes h_j+c\otimes d+
d\otimes c}\label{gR}
\end{eqnarray}
where $C_{\gamma+n\delta}\,~C_{(\delta-\gamma)+n\delta}\,,~\gamma\in\Delta^0
_+$ can be read off from (\ref{g2}) and the order in the product of (\ref{gR})
is defined by (\ref{ordering3}).
\vskip.1in
\noindent {\bf Remark:} $C_\gamma(q)$ have the
following general property
\begin{equation}
C_\gamma(q)=C_\gamma(q^{-1})~>~0\,~~~~{\rm for}~~q~>~0\,,~~ q~\neq~1
\label{positive}
\end{equation}
\vskip.1in
which follows directly from their definitions. The positivity (\ref{positive})
in fact implies useful information on unitarity of representations, which
we investigate elsewhere \cite{ZG}.
\section{Concluding Remarks}
In this Letter we have obtained the exlicit formulas of the universal
$R$-matrix for all quantized nontwisted rank 3 affine Lie algebras
$U_q(A_2^{(1)})\,,~~U_q(C_2^{(1)})$ and $U_q(G_2^{(1)})$ by deriving
all normalizing coeffients appearing the $R$-matrix.
These explicit expressions are
relevant with obtaining spectral parameter dependent solutions
to the Yang-Baxter equation for the corresponding quantum finite-dimensional
simple Lie algebras. The point
is that on finite-dimensional loop representations,
the infinite products in the expression of the universal $R$-matrix
truncate and thus are well defined. Since loop representations (say,
evaluation representations) automatically
contain a spectral parameter, the universal $R$-matrix, acting on
loop representations, automatically realizes the  so-called
"Yang-Baxterization" procedure
\cite{Jimbo2}\cite{ZGB2}. This is proved for the
rank 2 case $U_q(A_1^{(1)})$ in KT's paper\cite{KT2}. Therefore,
one may expect that our explicit expressions for the rank 3 case,
acting on loop representations, will truncate to the spectral parameter
dependent
solution to quantum Yang-Baxter equation for $U_q(A_2)\,,~~U_q(C_2)$
obtained in \cite{Jimbo2}\cite{ZGB2} and for $U_q(G_2)$ in \cite{Kuniba}.
\vskip.3in
\begin{center}
{\bf Acknowledgements:}
\end{center}
Y.Z.Z. would like to thank Anthony John Bracken for contineous encouragement
and suggestions, to thank Loriano Bonora for communication of preprint
\cite{KT2} and to thank M.Scheunert for many patient explanations on quantum
groups during July and August of last year. The financial support from
Australian Research Council
is gratefully acknowledged.

\end{document}